\documentclass[12pt,a4paper,twocolumn]{narms2}
\usepackage{amsmath}
\usepackage{graphicx}
\newcommand{\be}{\begin{equation}}
\newcommand{\ee}{\end{equation}}
\newcommand{\bi}{\begin{itemize}}
\newcommand{\ei}{\end{itemize}}
\newcommand{\im}{\item}

\usepackage{subfigure} 
\usepackage{psfig}
\usepackage{timesmt}   
\usepackage{chikako}   

\makeatletter
\renewcommand{\section}{\@startsection%
{section}%
{1}%
{0mm}%
{- \baselineskip}%
{0.15\baselineskip}%
{\normalfont\normalsize}}%

\renewcommand{\subsection}{\@startsection
{subsection}%
{2}%
{0mm}%
{-\baselineskip}%
{0.15\baselineskip}%
{\normalfont\normalsize}}%
\makeatother

\begin{document}

\title{Elasticity of sphere packings: pressure and initial state dependence}
\author{\large {I. Agnolin \& J.-N. Roux}\\
{\em Laboratoire des Mat\'eriaux et des Structures du G\'enie Civil, Institut Navier, 
Champs-sur-Marne, France}\\
}
\date{}
\abstract{ABSTRACT: Elastic properties and internal states of isotropic sphere packings are 
studied by numerical simulations. Several numerical protocols to assemble dense configurations 
are compared. One, which imitates experiments with lubricated contacts, produces well coordinated
states, while another, mimicking the effect of vibrations, results, for the same
density, in a much smaller coordination number $z$, as small as in much looser systems.
Upon varying the confining pressure $P$, simulations show a very nearly reversible variation of
density, while $z$ is irreversibly changed in a pressure cycle. Elastic moduli are shown to be mainly
related to the coordination number. Their $P$ dependence notably departs from
predictions of simple homogenization approaches in the case of the
shear moduli of poorly coordinated systems.}
\maketitle
\frenchspacing  
\section{INTRODUCTION}
The mechanical properties of solidlike granular materials are well
known to depend on the internal structure of the packing.
Classically, one distinguishes between the
behaviour of dense and loose
materials~\shortcite{DMWood,MIT93}. However, some results -- see e.g. 
\shortcite{Benahmed04} -- also indicate that other factors than the sole
packing fraction (or void index) also determine the quasistatic response to
applied load variations. In numerical simulations, it is a common
practice to remove friction in the assembling stage in order to
prepare dense samples~\shortcite{MGJS99,TH00}. It is not guaranteed
that the correct initial state, as obtained in laboratory experiments,
is reproduced. Elastic properties, or sound wave velocities, 
are now commonly measured in soil mechanics~\shortcite{CIJ88a,HI96} and
condensed matter physics~\shortcite{jia99} laboratories. Their evaluation in numerical
calculations can allow for comparisons with experiments.

We report here on a numerical study of isotropically assembled and
compressed sphere
packings, prepared in different initial states. Coordination numbers are found
to vary according to the preparation method independently from
density, and to determine the elastic moduli and their pressure
dependence. 
\section{MODEL AND NUMERICAL METHODS}
Numerical samples of 4000 identical balls of diameter $a$ and mass $m$
are prepared by standard
molecular dynamics calculations, involving periodic boundary
conditions in all three directions. The method is similar
to those of \citeN{CS79} or \citeN{TH00}, except that stresses, rather than strains,
are controlled, as in~\shortcite{PARA81}. Contact elasticity relates
the normal force
$F_N$ is to the normal deflection $h$ of
the contact by the Hertz law
\be
F_N = \frac{\tilde E\sqrt{a}}{3}h^{3/2}, \ \mbox{ with }\  \tilde E = \frac{E}{1-\nu^2},
\label{eqn:hertz}
\ee
while variations of tangential elastic forces ${\bf F}_T$ with
tangential displacements $\delta {\bf u}_T$,
\be
\frac{d{\bf F}_T}{d\delta {\bf u}_T} = \alpha_T
\frac{dF_N}{dh}, 
\ \mbox{ with } \  \alpha_T= \frac{2(1-\nu)}{2-\nu},
\label{eqn:mindlin}
\ee
are modeled with a simplified form of the Cattaneo-Mindlin-Deresiewicz
theory~\shortcite{JO85}. On implementing~\eqref{eqn:mindlin}, 
special care was taken, as advocated by~\citeN{EB96}, 
to avoid spurious creation of elastic energy.

Particles are endowed with the Young modulus $E=70$~GPa and the
Poisson coefficient $\nu=0.3$ of glass beads. The tangential reaction
is limited by the Coulomb condition with friction coefficient
$\mu=0.3$. 

First, the sample is assembled from an initial disordered loose granular gas
state, under the prescribed isotropic pressure $P=10$~kPa. In
procedure A, tangential forces are suppressed in this stage, as for
frictionless grains. This produces dense samples with a coordination
number $z^*$, counting only force-carrying grains and contacts,
approaching 6 in the rigid limit~\shortcite{JNR2000}. Only a small
fraction $f_0\simeq 1.3\%$ of grains (the ``rattlers'') carry no
force. This procedure, already used in other numerical
work~\shortcite{TH00}, amounts to dealing with perfectly lubricated
beads. Imperfect lubrication can be modelled with a very small
friction coefficient, $\mu_0 = 0.02$, resulting in slightly different
samples, denoted as B configurations.

A different assembling
procedure (method C) was designed to simulate dense samples obtained by
vibration. Configurations A are first dilated, scaling
all coordinates by a common factor $\lambda=1.005$, thereby supressing
the contacts ; then the grains are attributed random velocities 
and mixed with a kinetic-energy preserving event-driven (``hard
sphere'') calculation, until each of them has undergone 50 collisions
on average ; finally, they are compressed with strongly dissipative
collisions and friction, to mechanical equilibrium under $P=10$~kPa.
Remarkably (see table~\ref{tab:prep}), such configurations C
are very nearly as dense as the perfectly lubricated ones A, and
actually denser than B ones, while
their active coordination number $z^*$ is considerably lower, with
many more rattlers. 

Finally, procedure D consists in directly compressing the granular gas
to equilibrium at 10~kPa with the final coefficient of friction
$\mu=0.3$. $z^*$ values are
close to the C case, but the density is significantly lower.

Table~\ref{tab:prep} summarizes the data on these initial states. All
data throughout this paper are averaged over 5 different samples,
error bars correspond to one r.m.s. deviation.
\begin{table}[!htb]
\centering
\caption{Packing fraction $\Phi$, coordination number $z^*$ on force-carrying
  structure and proportion of rattlers $f_0$ at the lowest pressure
  $10$~kPa for the four simulated preparation procedures.
\label{tab:prep}
}
\small
\vskip 2mm
\begin{tabular}{|l||c|c|c|}  \cline{1-4}
State& $\Phi$ & $z^*$ & $f_0$ (\%)     \\
\hline
\hline
A&  $0.637\pm 0.009$ &$6.074\pm 0.002$ & $1.3\pm 0.2$\\
\hline
B&  $0.627\pm 2\cdot 10^{-4}$ &$5.80\pm 0.007$ & $1.65\pm 0.02$\\
\hline
C&$0.635\pm 0.002$ &$4.56\pm  0.03$  & $13.3 \pm 0.5$\\
\hline
D&$0.606\pm 0.002$ &$4.62\pm 0.01$ & $10.4\pm 0.9$\\
\hline
\end{tabular}
\vskip -2mm
\end{table}
These initial states are then further compressed, on applying pressure
steps, up to 100~MPa, assuming contacts still behave elastically,
and then the pressure is gradually decreased back to
10~kPa. To ensure quasistatic conditions are maintained with sufficient
accuracy, strain rates $\dot \epsilon$ are constrained by
condition $\dot \epsilon \sqrt{\frac{m}{aP}} <10^{-4}$. Equilibrium
states are recorded for pressure values at ratio 
$\sqrt{10}$. Throughout this process, the friction coefficient is
maintained equal to $0.3$, for all four configuration types.
Elastic constants are measured on building the stiffness matrix
associated with the contact network at equilibrium. 
\section{RESULTS}
\subsection{{\em Structure of equilibrium configurations}}
\begin{figure}[!htb]
\centering
\includegraphics[width=6.8cm,angle=270]{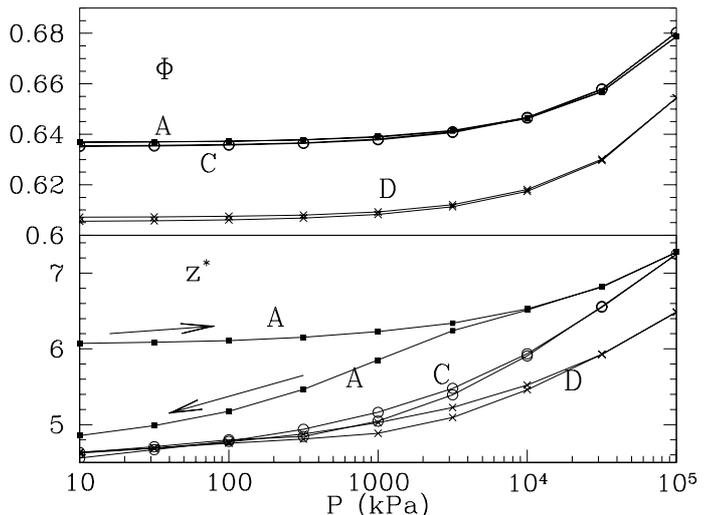}
\caption{Variations of $\Phi$ and $z^*$ as $P$ increases up to 100~MPa
  and decreases back to 10~kPa, in samples A (square dots), C
  (open circles), and D (crosses).
\label{fig:zphip}}
\end{figure}

The results obtained on samples A, C and D are reported below,
configurations of type B behaving very similarly to A ones.
Fig.~\ref{fig:zphip} displays the evolution of packing fraction $\Phi$
and coordination number $z^*$ over the compression-decompression
cycle. Changes in $\Phi$ are very nearly
reversible (elastic): density differences between states A, C and D,
are maintained at low pressure after one cycle, although A and C
samples exhibit very similar properties at high $P$. 
The shape of the normal force distribution also changes with $P$. It can be
characterized by the reduced moments:
\be
Z(\alpha) = \frac{\langle F_N^\alpha \rangle}{\langle
  F_N\rangle^\alpha},
\label{eqn:defz}
\ee
while the average normal force, for monosized beads, simply relates to
$P$ as 
\be
\langle F_N \rangle = \frac{\pi a^2 P}{z\Phi},
\label{eqn:fnp}
\ee
$z = z^*(1-f_0)$ being the total coordination number.
The width of this distribution, as expressed, e.g., by  $Z(2)$,
decreases at growing $P$, the fastest in well-coordinated A samples.
Changes in friction
mobilization are also observed~: as $P$ grows, it first decreases in C and D
samples, and increases in A ones (in which it starts at zero).

The change of $z^*$ in type A samples as $P$ decreases from a high
value -- many more contacts are lost than were gained at
increasing $P$ -- might seem surprising. One should note however
that configurations with
a high coordination number, for nearly rigid grains, are extremely
rare. Each contact requires a new equation to be satisfied by the set
of sphere centre positions. Equilibrium states of rigid,
frictionless sphere assemblies, apart from the motion of the scarce rattlers, are
isolated points in configuration space, because of isostaticity \shortcite{JNR2000}. As isotropic compression, at
the microscopic scale, is not reversible, due to friction
and to geometric changes, one should not expect exceptional
configurations to be retrieved upon
decreasing the pressure. Large coordination numbers of A (or B) samples
do not survive a pressure cycle. The
history of an isotropic sample can therefore 
significantly influence its structure without any appreciable 
density change. 
\subsection{{\em Elastic moduli}}
\begin{figure}[!htb]
\centering
\includegraphics[width=8.7cm]{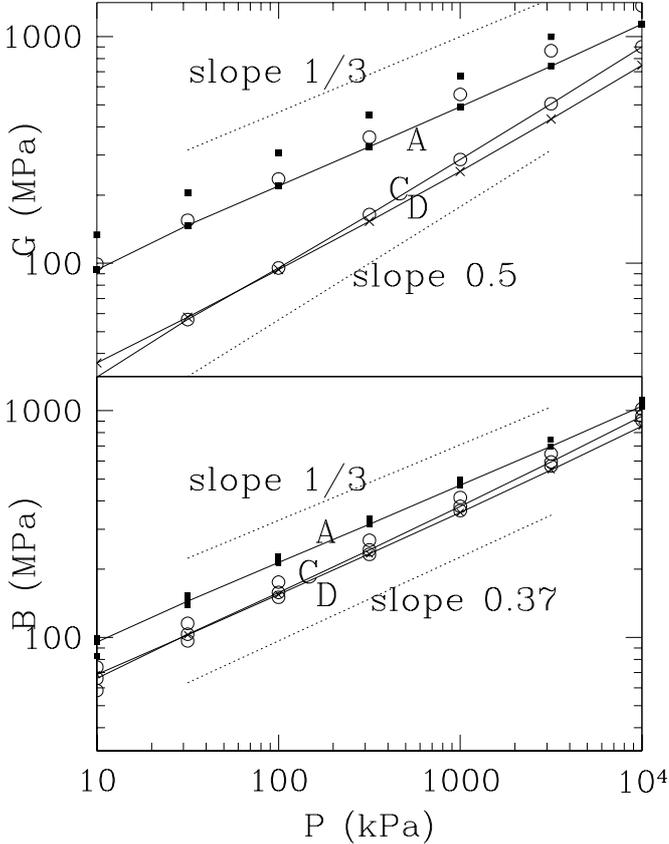}
\caption{$P$ dependence of bulk moduli $B$ (bottom) and
      shear moduli $G$ (top) (symbols for A, C, D states
as on fig.~\ref{fig:zphip}, joined by continuous lines), and of their
  upper ($B$ and $G$) and lower (for $B$ only) bounds for states A and
  C (same symbols, no line). Note the relatively narrow bracketing of
  $B$ in all cases, and the large overestimation of $G$ by
its Voigt upper bound in C samples.
\label{fig:modulp}
}
\end{figure}

Bulk ($B$) and shear ($G$) moduli of all equilibrium states for
ascending $P$ were computed on solving linear systems of equations
involving the stiffness matrix (they express the response to
infinitesimal stress changes). Their variations with $P$ are plotted
on Fig.~\ref{fig:modulp}.
Results for configurations B (not shown) are very close to those of A samples.
The obvious first conclusion to be drawn is that elastic moduli are
sensitive to coordination rather than density, as results for states $C$ and $D$
are very similar. The pressure dependence of bulk moduli differs a little
 between A samples on the one hand and C, D on the other. In the
 latter case, the increases of
$B$ with $P$ is slightly faster than the $P^{1/3}$ law predicted
by simple estimates (see below). The most striking behaviour is that
of $G$ in samples C and D, the increase of which approaches a
$P^{1/2}$ dependence. To explain such observations, one can try to estimate the
moduli as follows.  
Assuming the distribution of normal forces is known, one gets by
virtue of~\eqref{eqn:hertz} and~\eqref{eqn:mindlin} the distribution
of contact stiffnesses. It is easy, then, to derive upper bounds to
$B$ and $G$, and a lower bound to $B$, analogous to the Voigt and
Reuss bounds for elastic heterogeneous continua~\shortcite{NNH93}.
The Voigt upper bound is the simple ``effective medium'' estimate
that results from the assumption of affine displacement fields. Using the
properties and notations introduced in
Eqns.~\ref{eqn:hertz}, \ref{eqn:mindlin}, \ref{eqn:defz} and \ref{eqn:fnp},
one gets:
\be
\begin{aligned}
B&\le B^{\text{Voigt}}=\frac{1}{2}\left(\frac{z\Phi\tilde E}{3\pi}\right)^{2/3}P^{1/3}Z(1/3)\\
G&\le G^{\text{Voigt}}=\frac{6+9\alpha_T}{10} B^{\text{Voigt}}.
\end{aligned}
\label{eqn:voigt}
\ee
To write a lower bound (Reuss estimate), one needs a trial set of equilibrium
contact force increments corresponding to the stress increment.  For
a simple increase of isotropic pressure, this is readily obtained by a
scaling of the forces corresponding to the preexisting pressure. Hence
a lower bound for $B$ (but no such estimate is available for
$G$). Denoting as
$r_{TN}$ the ratio $\vert\vert {\bf F}_T\vert \vert
/F_N$ in each contact, and defining $$\tilde Z(5/3) = 
\frac{\langle F_N^{5/3}(1+\frac{r_{TN}^2}{\alpha_T})
\rangle }{\langle F_N\rangle ^{5/3}},$$
a modified reduced moment $Z(5/3)$ (Eqn.~\ref{eqn:defz}), one has:
\be
B\ge B^{\text{Reuss}}=\frac{1}{2}\left(\frac{z\Phi\tilde
  E}{3\pi}\right)^{2/3}
\frac{P^{1/3}}{\tilde Z(5/3)}.
\label{eqn:reuss}
\ee
In view of the force distribution and mobilization of friction
observed, $B$, bracketed by~\eqref{eqn:voigt} and~\ref{eqn:reuss},
cannot depart very much from a $P^{1/3}$ dependence
($r_{TN}^2/\alpha_T$ is at most $0.11$ anyway for $\mu=\nu=0.3$, and 
the product $Z(1/3)Z(5/3)$ only exceeds 1.15 for systems with few
contacts at low pressure). As to the increase of $z$ with $P$, it does
not appear to entail large effects either.

The behaviour of $G$ for samples C and D is quite different. $G$
seems to get unexpectedly small ($G< B/2$) at low pressure. Its
upper bound is a very poor estimate in such cases (as,
by~\eqref{eqn:voigt}, $G^{\text{Voigt}}= 1.34\times
  B^{\text{Voigt}}$). The situation is
reminiscent of frictionless sphere packings~\shortcite{OSLN03}, for
which $G\ll B$ under isotropic pressure in the rigid limit. We could
observe that those states had the largest level of
strain fluctuations (departure from affine displacement field). 

\section{CONCLUSION}
Our simulations of isotropically assembled sphere packings revealed
the following points.
\bi
\im
Configurations of a given density can vary considerably in
coordination number.
Samples assembled with a procedure designed to imitate vibration can
have a large density and a small coordination number.
\im
Elastic moduli are primarily sensitive to coordination numbers.
\im
They vary with pressure in rather good agreement to simple predictions
(nearly as $P^{1/3}$, with a small effect of contact creation as $P$
increases) in highly coordinated samples, but the shear modulus behaves quite
anomalously in poorly coordinated ones, with a low value at low $P$ and
a faster increase, nearly as $P^{1/2}$ in some cases.
\im
A compression-decompression cycle, although almost reversible in terms
of density, can substantially reduce the coordination number
when it was initially high.
\ei
We therefore suggest to use elastic moduli, which can be compared
between simulations and experiments, as indicators of the internal
state (contact density) of granular packings. 

On comparing numerically predicted ultrasonic wave velocities with
experimental values obtained on dense sphere packs with $P$ in the
range 100-800~kPa, we observe~\shortcite{Xiaoping-ici} that perfectly
lubricated samples (type A or B), are considerably too stiff, even though
they agree with experimental observations in the MPa range~\shortcite{MGJS99}. Although
somewhat idealized, our ``vibrated'' ones (type C) seem to be closer 
to the materials studied in the laboratory. We also note in another
contribution to the present proceedings \shortcite{JN-ici} that their stress-strain curves
under growing deviatoric stress are also closer to
experimentally observed mechanical behaviours.
Of course, it will be necessary in the near future to investigate  by
numerical simulations more ``realistic'' assembling
procedures~\shortcite{Sacha-ici}, and the effects of the resulting anisotropy.

\bibliographystyle{chikako}      

\end{document}